\documentclass[twocolumn,showpacs,preprintnumbers,nofootinbib,prd,superscriptaddress,groupedaddress,10pt]{revtex4-2}
\usepackage{graphicx,amssymb,amsmath,amsthm,amsfonts,epsfig}
\usepackage[linktocpage]{hyperref}
\usepackage[usenames]{color}
\usepackage{epstopdf}

\usepackage{bm}
\usepackage{dcolumn}
\usepackage[latin1]{inputenc}
\usepackage{latexsym}
\usepackage{rotating}
\usepackage{hyperref}
\usepackage{color}
\usepackage{longtable}
\usepackage{enumerate}
\usepackage{mathtools}
\usepackage{url}
\newcommand{\ben}{\begin{enumerate}}
\newcommand{\een}{\end{enumerate}}

\def\be{\begin{equation}}
\def\ee{\end{equation}}
\def\bea{\begin{eqnarray}}
\def\eea{\end{eqnarray}}
\newcommand{\beq}{\begin{eqnarray}}
\newcommand{\eeq}{\end{eqnarray}} 
\newcommand{\ba}{\begin{align}}
\newcommand{\ea}{\end{align}}

\def\ba{\bar{a}}

\newcommand{\tn}{\textnormal}

\begin{document}

\title{Black holes in an Effective Field Theory extension of GR}
\author{
Vitor Cardoso$^{1,2}$,
Masashi Kimura$^{1}$,
Andrea Maselli$^{1}$,
Leonardo Senatore$^{3}$
}
\affiliation{${^1}$ CENTRA, Departamento de F\'{\i}sica, Instituto Superior T\'ecnico -- IST, Universidade de Lisboa -- UL,
Avenida Rovisco Pais 1, 1049 Lisboa, Portugal}
\affiliation{${^2}$ CERN 1 Esplanade des Particules, Geneva 23, CH-1211, Switzerland}
\affiliation{${^3}$ SITP and KIPAC, Department of Physics and SLAC, Stanford University, Stanford, CA 94305}
\begin{abstract}
Effective field theory methods suggest that some rather-general extensions of General Relativity include, or are mimicked by, certain
higher-order curvature corrections, with coupling constants expected to be small but otherwise arbitrary.
Thus, the tantalizing prospect to test the fundamental nature of gravity with gravitational-wave observations, in a systematic way, emerges naturally.
Here, we build black hole solutions in such a framework and study their main properties. 
Once rotation is included, we find the first purely gravitational example of geometries without $\mathbb{Z}_2$-symmetry.
Despite the higher-order operators of the theory, we show that linearized fluctuations of such geometries obey second-order differential equations.
We find nonzero tidal Love numbers. We study and compute the quasinormal modes of such geometries. These results are of interest to gravitational-wave science but also potentially relevant for electromagnetic observations of the galactic center or $X$-ray binaries.
\end{abstract}
\maketitle

\noindent{\bf{\em Introduction.}}
The gravitational-wave (GW) astronomy era has began. As the sensitivity of our GW detectors increases,
so does our ability to realize the potential of the field. From a mapping of compact objects throughout
the visible universe, to measurements of the cosmological expansion rate, the opportunities in both astrophysics and fundamental physics are numerous~\cite{Barack:2018yly}.
GWs carry information from regions of spacetime where gravity is ``strong'', and therefore are genuine probes of one of the most
surprising predictions of General Relativity (GR): black holes (BHs). In GR, the uniqueness theorems suggest that BHs are all well described by the Kerr geometry. Such geometry is fully determined by two parameters (mass and spin) which also characterize the full multipolar structure of these objects~\cite{Berti:2015itd,Cardoso:2016ryw,Herdeiro:2015waa,Sotiriou:2015pka,Barack:2018yly}. 
A simple, well-known method to test such simple multipolar structure is by measuring accurately the final stages of relaxation of a perturbed BH, such as the one created after the inspiral and merger of a BH binary~\cite{Berti:2005ys,Berti:2016lat,Yang:2017zxs}.

There are compelling reasons to test the nature and the geometry of compact objects such as BHs~\cite{Cardoso:2017cqb,Cardoso:2017njb}. 
In GR, BHs harbour singularities where the gravitational field becomes unbounded and where quantum effects may be screened from outside observers thanks to the event horizon. The assertion that all singularities are hidden from us is so remarkable that any observational evidence for or against it is welcome. Furthermore, the complications associated to putting together GR and quantum mechanics near the Planck scale have been challenging thus far. One resolution is provided by string theory, 
which is nevertheless a remarkable extension to our laws of physics.
 
It is thus an extremely appealing prospect that GW astronomy may be able to improve, even if only incrementally, our understanding of the nature of gravity at high energies.

Technically, in the modern way in which physical laws are written, gravity is ``extended'' by adding to the GR Lagrangian either new massive particles, or new terms built out of the Riemann tensor, suppressed by some mass scale. For associated mass scales much larger than a few inverse km (the scale probed by current GW observations), it is very hard for these observations to provide any observational guidance on the nature of gravity. Thus, there are theoretical prejudices against embarking upon
missions set to test extensions of GR at current scales. However, such preconceptions need not be true: extensions of GR at scales that are probed by experiments such as LIGO/VIRGO are theoretically possible, and not excluded by current experiments or observations. They should be probed.

Effective Field Theory (EFT) is a useful guide in the search for the most general extension to GR, under the following assumptions: the theory should be testable with GW observations; should be consistent with other experiments, including short distance tests of GR; should agree with widely accepted principles of physics, such as locality, causality and unitarity; it should not involve new light degrees of freedom. Such an EFT, which is unique, was constructed recently, and has appealing features~\cite{Endlich:2017tqa}. By studying the signatures of one single Lagrangian, a vast class of theories is covered: indeed it is guaranteed that any theory satisfying the above-mentioned assumptions will lead to the same signatures as the EFT (for {\it some} choice of parameters). However, by using EFT, all information is summarized in constraints (or measurements, if one is lucky) of very few coefficients. Finally, in contrast to phenomenological parametrizations, the one offered by EFTs automatically constrains the observational investigation to the space of physically-viable theories, and, within this class, offers the optimal general parametrization of the observational signals.

\noindent{\bf{\em Effective theory.}}
Consider the effective action~\cite{Endlich:2017tqa}
\beq
S_{\rm eff} &=& \int d^4 x \sqrt{-g}\,2 M_{\rm pl}^2 \left(
R - \frac{{\cal C}^2}{\Lambda^6} - \frac{{\cal \tilde{C}}^2}{\tilde{\Lambda}^6} - \frac{{\cal \tilde{C}{\cal C}}}{\Lambda_-^6} \right),\label{eq:eftaction}\\
{\cal C} &\equiv& R_{\alpha \beta \gamma \delta} R^{\alpha \beta \gamma \delta}\,,\quad {\cal \tilde{C}} \equiv R_{\alpha \beta \gamma \delta} \tilde{R}^{\alpha \beta \gamma \delta}\,,\nonumber
\eeq
where the dual tensor
$\tilde{R}^{\alpha \beta \gamma \delta}\equiv\epsilon^{\alpha \beta}{}_{\mu \nu} R^{\mu \nu \gamma \delta}$, 
and $\epsilon_{0123} = \sqrt{-g}$.
The equations of motion, for $R_{\mu\nu}=0$, read:
\beq
&&R^{\mu \alpha} - \frac{1}{2}g^{\mu \alpha} R = \frac{1}{\Lambda^6}\left(8 R^{\mu \nu \alpha \beta} \nabla_\nu \nabla_\beta {\cal C}
+ \frac{g^{\mu \alpha}}{2}{\cal C}^2\right)\nonumber\\
&+&\frac{1}{\tilde{\Lambda}^6}\left(8 \tilde{R}^{\mu \rho \alpha \nu}\nabla_\rho \nabla_\nu {\cal \tilde{C}}+\frac{1}{2}g^{\mu \alpha}{\cal \tilde{C}}^2
\right)\nonumber\\
&+& \frac{1}{\Lambda_-^6} \left(4\tilde{R}^{\mu \rho \alpha \nu} \nabla_\rho \nabla_\nu {\cal C}
+4 R^{\mu \rho \alpha \nu} \nabla_{\rho}\nabla_{\nu}{\cal \tilde{C}}+\frac{g^{\mu \alpha}}{2}{\cal \tilde{C}{\cal C}}
\right)\,,\label{eq:effectivefieldeq}
\eeq
and hold around a vacuum solution at order ${\cal O}(1/\Lambda^6, 1/\tilde{\Lambda}^6, 1/\Lambda_-^6)$.

\noindent{\bf{\em Spherically symmetric solutions.}}
From now on, we use dimensionless coupling parameters
\be
\left(\epsilon_1,\epsilon_2,\epsilon_3\right)=\left(\frac{1}{M^6\Lambda^6},\frac{1}{M^6\tilde{\Lambda}^6},\frac{1}{M^6\Lambda_-^6}\right)\,,
\ee
where $M$ is the gravitational mass of the spacetime~\cite{EFT}.
Consider spherically symmetric, static vacuum solutions
\be
ds^2 = - f_t^{\epsilon_i}(r) dt^2 + \frac{1}{f_r^{\epsilon_i}(r)}dr^2 + r^2 (d\theta^2 + \sin^2\theta d\phi^2)\,.\label{eq:sphericalsol}
\ee
In this setup, ${\cal \tilde{C}} = 0, \ \tilde{R}^{\mu \rho \alpha \nu} \nabla_\rho \nabla_\nu {\cal C} = 0$. Since these are small corrections to GR, we look for slight deviations from the Schwarzschild geometry, $f_{t,r}^{\epsilon_i}=1-2M/r+\epsilon_i\delta f_{t,r}^{\epsilon_i}$. Asymptotic flatness yields~\cite{units}
\beq
f_t^{\epsilon_i} &=& 1 - \frac{2M}{r} + \epsilon_i \delta^i_1 \left(-\frac{1024 M^9}{r^9}+\frac{1408 M^{10}}{r^{10}}\right)\,,\label{gtt1}\\
f_r^{\epsilon_i} &=& 1 - \frac{2M}{r} + \epsilon_i \delta^i_1 \left(-\frac{4608 M^9}{r^9}+\frac{8576 M^{10}}{r^{10}}\right)\,.\label{grr2}
\eeq
This spacetime describes a BH, with an event horizon located at $r=r_H = 2 M \left(1 +5\epsilon_1/8\right)+{\cal O}(\epsilon_i^{2})$.
As expected (these are higher-curvature modifications) the corrections decay very fast with the radial distance $r$. The usual PPN parameters $\gamma, \beta$ for example~\cite{Will:2005va},
are the same as GR. Likewise, GWs in this theory propagate at the speed of light when far from sources and the dispersion
relation is the same as GR. ${\cal{O}}(\epsilon^2_1)$ corrections are shown in the supplemental material, indicating that these solutions can be trusted for $\epsilon_1\lesssim 0.04$. However, one should keep in mind that the leading corrections are of order $\epsilon^{4/3}\sim{\cal{O}}(1/\Lambda^8)$ from operators schematically of order $R_{\mu\nu\rho\sigma}{}^5$ in~(\ref{eq:eftaction})~\cite{Endlich:2017tqa}.

\noindent{\bf{\em Slowly-spinning BHs and $\mathbb{Z}_2$-symmetry breaking.}}
The geometry above can be extended to include angular momentum as:
\beq
ds^2&=&-f_t[1+2h]dt^2+[1+2m]\frac{dr^2}{f_r}\nonumber\\
&+&r^2[1+2k]\Big[d\theta^2+\sin^2\theta(d\phi-\hat{\omega} dt)^2\Big]\ ,
\eeq
where $f_t, f_r$ are the non-spinning metric functions \eqref{gtt1}-\eqref{grr2}, and 
$(h,m,k,\hat{\omega})$ are functions of $r, \theta$ only, and perturbative in the angular momentum $J=M^2\chi$. 
The angular dependence can be expanded in terms of the Legendre polynomials, while the radial components are expressed as a series in powers 
of $\chi$~\cite{Hartle:1967he,Hartle:1968si}.
Inserting the previous ansatz into the field equations \eqref{eq:effectivefieldeq} we can solve for the unknown 
metric functions ($h,m,k,\hat{\omega}$) order by order in the BH spin~\cite{Maselli:2015tta}. 
The full solution was computed up to $\mathcal{O}(\chi^4)$. The explicit form 
of the metric tensor and other quantities in this paper can be found online~\cite{Cardosoweb}.

Having derived the full metric it is straightforward to compute the main properties of the slowly spinning BH 
at the desired order in $J$. The horizon, for example, is located at
\be
\frac{r_{H}}{M}=2+\frac{5\epsilon_1}{4}-\chi ^2 \left(\frac{1}{4}+\frac{51\epsilon_1}{352}-\frac{72\epsilon_2}{11}\right)\,.
\ee
Note that $r_H$ depends on even powers of $\chi$ only. Moreover, the spin terms turn on $\epsilon_2$-corrections 
which were absent for the static solutions. The equatorial frequency at the ISCO and at the photosphere are
\beq
M\Omega_\tn{ISCO}&=&\frac{1}{6\sqrt{6}}-\frac{5291\epsilon_1}{1889568 \sqrt{6}}\nonumber\\
&+&\chi\left(\frac{11}{216}-\frac{2059367\epsilon_1}{374134464}+\frac{211\epsilon_2}{144342}\right)\,,\label{omegaI}\\
%
%
%
M\Omega_\tn{LR}&=&\frac{1}{3 \sqrt{3}}-\frac{832\epsilon_1}{59049 \sqrt{3}}\nonumber\\
&+&\chi\left(\frac{2}{27}-\frac{159872\epsilon_1}{5845851}+\frac{2048\epsilon_2}{72171}\right)\,.
%
\eeq
Even though $\epsilon_1\sim 1$ induces changes in the horizon area of order ${\cal O}(1)$, this is not an observable. By contrast,
the gauge-independent frequency $\Omega_{\rm ISCO}$ suffers much smaller corrections even for such large couplings: a small increase in radial distance translates into a region of slightly smaller curvature and much less affected by such curvature-sensitive terms.
This is also true for other relativistic effects associated to frame dragging. For example, the orbital plane of 
test particles, if not aligned with the equatorial plane, will precess around the angular momentum axis of the rotating body, the Lense-Thirring effect~\cite{Mashhoon:1984hi}. We can investigate this phenomenon studying the precession frequencies $\Omega_r$ and $\Omega_{\theta}$, which describe the perturbation in circular orbits due $r$ and $\theta$ velocity components~\cite{Stella:1998jl,Maselli:2017kic} (analytic expressions for these quantities can be found online~\cite{Cardosoweb}). 

The BH quadrupole moment is~\cite{Thorne:1980jl}
\be
Q_{20}=-\sqrt{\frac{64\pi}{15}}\chi^2\left(1-\frac{89\epsilon_1}{25}+\frac{216\epsilon_2}{25}\right)\,.
\ee

Observations probing only the background geometry, such as those using electromagnetic observations of matter close to the galactic center can be used to constraint the coupling parameters $\epsilon_i$. Most of the constraints on alternatives to the Kerr geometry use ad hoc parametrizations~\cite{Berti:2015itd,Barack:2018yly}. The geometry above either does not fit into such parametrized metrics, or when it does, was never studied in the context of actual observations (see e.g. Refs~\cite{Johannsen:2015hib,Johannsen:2016vqy}).

The $\epsilon_3$ parameter adds an interesting ``twist'' to the solutions, not apparent from equatorial observables. At linear order in rotation, such coupling introduces the following corrections to the background geometry
\begin{align}
\delta g_{\mu\nu}dx^\mu dx^\nu& =\epsilon_3 \chi  \bigg(\frac{73728M^9}{r^9}dr^2 
+\frac{256M^9(243M-160r)}{5r^8}
\nonumber\\ & \times
(d\theta^2 + \sin^2\theta d\phi^2)\bigg)\cos\theta\,.
\end{align}
This is the first purely gravitational example of a $\mathbb{Z}_2$-symmetry violating BH solution~\footnote{In the last stages of preparation of this work, Ref.~\cite{Cunha:2018uzc} reports a similar finding, but for theories with non-minimal couplings and extra fields.}. The above form is not a coordinate artifact: curvature scalars at the horizon are indeed affected by it, for example $R=27\chi\epsilon_3\cos\theta/(4M^2),\,\delta {\cal C}=189 \chi \epsilon_3\cos\theta/(8M^4),\,\delta \tilde{\cal C}=0$, to linear order in spin. 
The linear stability of this solution is unclear at this stage.

\noindent{\bf{\em Dynamics.}}
Consider now the dynamics of the spherically symmetric solution~\eqref{eq:sphericalsol}. Fourier-decompose and expand the fluctuations in spherical tensor harmonics.
The tensor harmonics are of odd (``$-$'') and even-type (``$+$''), depending on their transformation properties, and these different sectors {\it usually} decouple~\cite{Regge:1957td,Zerilli:1971wd,Berti:2009kk}. The form of the perturbed metric is 
\be
h^{\rm (-)}_{\mu \nu}dx^\mu dx^\nu =2e^{-i\omega t}  \sin\theta \partial_\theta Y_{\ell 0}d\phi \left(h_0 dt+h_1 dr \right)\,,\nonumber
\ee
and
\beq
h^{\rm (+)}_{\mu \nu}dx^\mu dx^\nu &=& e^{-i\omega t} Y_{\ell 0}\Big{(}f_t H_0 dt^2 + 2 H_1 dtdr 
\nonumber\\
&+&f_r^{-1}H_2 dr^2 + r^2 K (d\theta^2 + \sin^2\theta d\phi^2)\Big{)}\,,\nonumber
\eeq
respectively, where $h_0, h_1, H_0, H_1, H_2$ and $K$ are functions of $r$, 
and $Y_{\ell 0}$ is the spherical harmonics $Y_{\ell m=0}$.

Given the structure of the field equations, the equations of motion are described by up-to fourth-order differential equations.
However, i) when the coupling parameters $\epsilon_i$ are taken to be perturbatively small, one can use the zeroth-order GR equations to reduce the problem to second order differential equations;
ii) the odd and even parity perturbations around the spacetime Eq.~\eqref{eq:sphericalsol} 
for the theory Eq.~\eqref{eq:eftaction} are decoupled for $\epsilon_1$ and $\epsilon_2$ corrections
but coupled for $\epsilon_3$ corrections, due to the parity violating terms.

To derive the perturbed field equations
we replace the metric $g_{\mu \nu} + h^{\rm (-)}_{\mu \nu} +  h^{\rm (+)}_{\mu \nu}$ into 
Eq.~\eqref{eq:effectivefieldeq} and expand the equations 
up to ${\cal O}(h_{\mu \nu}^{(\pm)})$ and ${\cal O}(\epsilon_i)$. 
In GR, the perturbed field equations around the Schwarzschild spacetime can be combined into single master functions,
\beq
&& \Psi^{\rm GR}_- = \frac{i f h_1(r)}{r \omega }\,,\nonumber
\\
&& \Psi^{\rm GR}_+= \frac{1}{\lambda r+6 M}\left[-r^2 K+\frac{i f r H_1}{\omega }\right]\,,\nonumber
\eeq
and they satisfy the Regge-Wheeler and Zerilli equations,
\be
\frac{d^2\Psi_\pm^{\rm GR}}{dr_*^2}
+\left(\omega^2-f V_\pm^{\rm GR}\right)\Psi_\pm^{\rm GR}=0\,,
\label{eq:RWeqschwarzschild}
\ee
with
\beq
V_-^{\rm GR} &=& \frac{(\lambda+2)}{r^2}-\frac{6M}{r^3}\,,
\\
V_+^{\rm GR}  &=&\frac{1}{r^3 (\lambda r+6 M)^2}\times\Big[36\lambda\,M^2\,r+6\lambda^2Mr^2 \nonumber\\ 
& +&  \lambda^2(\lambda+2) r^3+72 M^3\Big]\,,
\eeq
where $\lambda=\ell^2+\ell-2$, $f =1 - 2M/r$ and $dr/dr_* = f$. All components of $h_{\mu \nu}^{(\pm)}$, i.e., $h_0, h_1, H_0, H_1, H_2$ and $K$, 
are expressed by master variables $\Psi_\pm^{\rm GR}$ at this lowest order.

Clearly, $\Psi_\pm^{\epsilon_i}=\Psi_\pm^{\rm GR}+{\cal O}(\epsilon_i)$.
At linear order in $h_{\mu \nu}^{(\pm)}$, the right-hand-side of Eq.~\eqref{eq:effectivefieldeq}
takes the form $\epsilon_i \times {\cal O}(h_{\mu \nu}^{(\pm)})$.
We can use the ${\cal O}(\epsilon_i^0)$ relations among $h_{\mu \nu}^{(\pm)}$ and the master variables 
$\Psi_\pm^{\epsilon_i}$ to compute ${\cal O}(h_{\mu \nu}^{(\pm)})$ terms 
in the right-hand-side of Eq.~\eqref{eq:effectivefieldeq},
because ${\cal O}(\epsilon_i)$ corrections in $h_{\mu \nu}^{(\pm)}$ become higher order in $\epsilon_i$.
In this way, all the $h_{\mu \nu}^{(\pm)}$ in the right-hand-side of Eq.~\eqref{eq:effectivefieldeq} 
can be replaced with $\Psi_\pm^{\epsilon_i}$.
At ${\cal O}(\epsilon_i^0)$, $\Psi_\pm^{\epsilon_i}$ satisfies Eq.~\eqref{eq:RWeqschwarzschild},
hence by using Eq.~\eqref{eq:RWeqschwarzschild} and its derivatives with respect to $r$, 
we can replace higher order derivatives into lower derivatives. The above is a procedure to obtain second-order equations of motion at linear level. For an initial study of how to do this at non-linear level, see for example Ref.~\cite{Allwright:2018rut}.

\noindent{\bf{\em Tidal Love numbers.}}
When an object is placed in an external gravitational field, it is tidally deformed.
Such effect can be quantified through the calculation of the object's Tidal Love Numbers (TLNs)~\cite{Flanagan:2007ix,Binnington:2009bb,Damour:2009vw}.
BHs in GR possess the remarkable property that their TLNs are zero~\cite{Binnington:2009bb,Damour:2009vw,Porto:2016pyg,Porto:2016zng,Rothstein:simons,Cardoso:2017cfl}.
Non-zero TLNs are therefore a good discriminator of new physics~\cite{Kol:2011vg, Cardoso:2017cfl,Maselli:2017cmm}.
We find the following quadrupolar TLNs for BHs in this theory, using the conventions of Ref.~\cite{Cardoso:2017cfl},
\beq
k_2^{E}&=&-\frac{1008}{25}\epsilon_1\,,\quad k_2^{B}=\frac{432}{25}\epsilon_1\,,\\
k_2^{E}&=&0\,,\quad \qquad k_2^{B}=-\frac{384}{5}\epsilon_2\,,
\eeq
for polar and axial-type quadrupolar external fields, respectively.
For $\epsilon_3$ corrections, we find that an external {\it axial} quadrupolar field, induces a {\it polar} quadrupolar moment in the spacetime.
New TLNs seem to be needed to describe this scenario, but we will not dwell on this any further.

\noindent{\bf{\em Quasinormal modes and stability.}}
After the procedure above, the dynamics of slightly disturbed non-spinning BHs  
in this theory are described by very simple master equations, presented in the online \texttt{MATHEMATICA} notebook~\cite{Cardosoweb}. 
It should be noted that the master equations are of second order in spatial and time derivatives, with no mixing. Higher-order terms cancel out.
The master equations are completely decoupled for $\epsilon_1, \epsilon_2$.
Computing the characteristic quasinormal frequencies is then a straightforward procedure~\cite{Berti:2009kk,Cardosoweb}.
Define, for each multipole $\ell$,
\be
\delta_l\equiv \left(\frac{{\rm Re}(\omega-\omega_0)}{\epsilon_i{\rm Re}(\omega_0)}\,,\frac{{\rm Im}(\omega-\omega_0)}{\epsilon_i{\rm Im}(\omega_0)}\right)\,,
\ee
where $\omega_0$ is the unperturbed, GR value.

\noindent{\em $\epsilon_1$ corrections.}
We used a direct integration approach to compute the quasinormal modes of odd- and even parity perturbations, and two different independent codes.
We find
\beq
\delta^{\rm even}_2&=&(0.45, -2.75)\,,\quad\delta^{\rm odd}_2=(0.22, -0.64)\,,\nonumber\\
\delta^{\rm even}_3&=&(1.07, -6.42)\,,\quad\delta^{\rm odd}_3=-(0.0099, 0.44)\,,\nonumber\\
\delta^{\rm even}_4&=&(1.76, -11.43)\,,\quad\delta^{\rm odd}_4=-(0.048, 0.19)\,,
\eeq
%

\noindent{{\em $\epsilon_2$ corrections.}}
For $\epsilon_2$ couplings, we find
\beq
\delta^{\rm even}_2&=&0\,,\quad \delta^{\rm odd}_2=(2.18, -15.85)\,,\nonumber\\
\delta^{\rm even}_3&=&0\,,\quad\delta^{\rm odd}_3=(4.23, -30.47)\,,\nonumber\\
\delta^{\rm even}_4&=&0\,,\quad\delta^{\rm odd}_4=(6.99, -49.54)\,.
\eeq
Note that ${\cal \tilde{C}} = {\cal O}[(h_{\mu \nu}^{(+)})^2]$ for even perturbations which explains why its even sector is 
identical to GR. Notice also that even (odd) corrections scale as $\sim l^2$ for $\epsilon_1$ ($\epsilon_2$) couplings. In fact, at large $l$ there is an extrema in the potential at $r=11M/5$. Thus, these modes are localized away from the light ring and closer to the horizon. This introduces a new scale in the problem, and raises the interesting possibility that
higher multipoles arrive later, producing a phenomena {\it similar} to echoes in the late-time
waveform~\cite{Cardoso:2016rao,Abedi:2016hgu,Cardoso:2017cqb,Mark:2017dnq,Cardoso:2017njb,Correia:2018apm}. 
The results for $\epsilon_3$ corrections will be discussed elsewhere. At sufficiently small couplings (including most likely the regime of validity of the EFT), the 
quasinormal frequencies are only small corrections to the GR values, and the system is therefore stable.
Notice that from a purely mathematical point of view,  the structure of the master equations is such that the system is linearly mode-stable for any positive $\epsilon_i$
(see supplemental material).

Constraints on this theory can be obtained through a careful analysis of the post-Newtonian regime~\cite{Endlich:2017tqa}. However, due to the strong-dependence on the curvature, probes of the region close to the horizon - such as quasinormal modes - are clearly at an advantage. From the requirement that $(\omega-\omega_0)/\omega_0\lesssim 1$, we find that LIGO observations constrain $(1/\Lambda,1/\tilde{\Lambda},1/\Lambda_-)\sim 100 {\rm Km}$, from the first event GW150914~\cite{TheLIGOScientific:2016src}. One can foresee improvements of one order of magnitude, but not more than that, from ringdown signals. The space-based detector LISA will be sensitive to lower frequencies, therefore larger masses, and so different $\Lambda$'s~\cite{Audley:2017drz} (see~\cite{Endlich:2017tqa} for a discussion).

\noindent{\bf{\em Discussion.}}
The search for signatures of deviations from Einstein's gravity is a challenging program. Theoretical bias may lead us to expect no changes at all, on scales which current detectors can probe. However, if one abandons such prejudice, the possibilities to change gravity are very large. Thus, it is customary to focus on possible generic smoking-gun effects, or then on a handful of theories which one knows well enough to calculate observables~\cite{Barack:2018yly,Berti:2015itd}. One appealing simple and generic class of theories consists on either minimally or weakly coupled light scalars~\cite{Arvanitaki:2009fg}. Such a broad class of theories gives rise to clear signatures, observable in both GW detectors and in traditional telescopes~\cite{Brito:2014wla,Arvanitaki:2016qwi,Brito:2017zvb,Ferreira:2017pth,Boskovic:2018rub}. In the absence of scalar degrees of freedom, EFT arguments suggest that Eq.~\ref{eq:eftaction} describes the dominant effects of the most generic class of theories compatible with reasonable requirements~\cite{Endlich:2017tqa}. 

Our result for the curvature corrected geometry of non-spinning geometries, Eqs.~\eqref{gtt1}-\eqref{grr2}, shows why it is difficult to probe such corrections: they decay extremely fast at large distances, and even at the light ring they are suppressed (in fact, such result generalizes easily to higher-order theories, 
as shown in the Supplemental material). The observational prospects appear to be much more promising for the signal associated to the quasi normal modes.
%

The results we derived are of interest for GW detectors, but also to VLBI observatories such as GRAVITY~\cite{Abuter:2018drb} or the Event Horizon telescope~\cite{Psaltis:2018xkc,Falcke:2013ola}, providing a physically motivated and understood geometry with which to perform tests of GR.

\begin{acknowledgments}
L.S. would like to thank Junwu Huang and Victor Gorbenko for collaboration in the very early stages of this work.
We would like to thank Leonardo Gualtieri for useful comments on an earlier version of this manuscript, and Takuya Katagiri for pointing our typos in the original version of Eq. (16).
V.C. and L.S. would like to thank ICTP for hospitality where part of the work was initiated during the ``Summer School on Cosmology 2018.'' 
The authors acknowledge financial support provided under the European Union's H2020 ERC 
Consolidator Grant ``Matter and strong-field gravity: New frontiers in Einstein's theory'' grant 
agreement no. MaGRaTh--646597. 
This project has received funding from the European Union's Horizon 2020 research and innovation programme under the Marie Sklodowska-Curie grant agreement No 690904.
The authors would like to acknowledge networking support by the GWverse COST Action CA16104, ``Black holes, gravitational waves and fundamental physics.''
L.S. is partially supported by Simons Foundation Origins of the Universe program
(Modern Inflationary Cosmology collaboration) and by NSF award 1720397.
Computations were performed on the ``Baltasar Sete-Sois'' cluster at IST.
\end{acknowledgments}

\appendix 

\section{The master equations}

\subsection{$\epsilon_1$ corrections}
For $\epsilon_2=\epsilon_3=0$,  the perturbed field equations take the following form,
\be
\frac{d^2\Psi_\pm^{\epsilon_1}}{dr_*^2}+\left(\omega^2-\sqrt{f_t^{\epsilon_1}f_r^{\epsilon_1}}\left(V_\pm^{\rm GR}
+\epsilon_1 V_\pm^{\epsilon_1}\right)\right)\Psi_\pm^{\epsilon_1} =0\,,\label{eq:epsilon1mastereq}
\ee
where $dr/dr_*=\sqrt{f_t^{\epsilon_i}f_r^{\epsilon_i}}$. We should stress that these equations are valid only up to $O(\epsilon_1)$.
%
For odd parity perturbations, the master variable and potential are
\beq
\Psi_-^{\epsilon_1} &=& \frac{i \sqrt{f_t^{\epsilon_1}f_r^{\epsilon_1}} h_1}{\omega r} \left(1 + \frac{1152 M^8 \epsilon_1 (13 M - 7 r)}{r^9}\right),\\
V_-^{\epsilon_1} &=& -\frac{256M^8}{r^{12}}\times \Big(15561M^2+Mr(146\ell(\ell+1)\nonumber\\
&-& 13509) +9r^2\left(324-8\ell(\ell+1)+7r^2\omega^2\right)\Big)\,.
\eeq
For even parity perturbations, 
\beq
 \Psi_+^{\epsilon_1} &=&\frac{i \sqrt{f_t^{\epsilon_1}f_r^{\epsilon_1}} H_1}{\omega r^8 ((\ell (\ell+1)-2) r+6 M)^2}\nonumber\\ 
&\times &\Big[r^9 ((\ell (\ell+1)-2) r +6 M)\nonumber\\
&-&128 M^8 \epsilon_1 \Big((23 \ell (\ell+1)  +98) M r\nonumber\\
&+& 3 (\ell-1) \ell (\ell+1) (\ell+2) r^2-162M^2\Big)\Big]\nonumber\\
&+&  \frac{K}{r^7 ((\ell (\ell+1)-2) r+6 M)^2}\nonumber\\ &\times & \Big[r^9 ((\ell (\ell+1)-2) r+6 M) \nonumber\\
&-& 384 M^8\epsilon_1 \Big(3(\ell^2+\ell+14) M r\nonumber\\ 
&+& (\ell-1) \ell (\ell+1) (\ell+2) r^2 -82 M^2\Big)\Big].\\
V_+^{\epsilon_1}  &=& \frac{128 M^8}{r^{12} ((\ell^2+\ell-2) r+6 M)^3}\nonumber\\ 
&\times & \Big[-126 r^4 \omega ^2 ((\ell^2+\ell-2) r+6 M)^3\nonumber\\
&+& 230688 M^5 + 144 (1984 \ell (\ell+1)-4643) M^4 r\nonumber\\
&+&36 (\ell-1) (\ell+2) (2249 \ell (\ell+1)-10132) M^3 r^2\nonumber\\
&+&18 (\ell-1) (\ell+2) (\ell (\ell+1) (321 \ell (\ell+1)-4982)\nonumber\\
&+& 10696) M^2 r^3-10 (\ell^2+\ell-2)^2 (\ell (\ell+1) (13 \ell (\ell+1)\nonumber\\
&+& 469)-2502) M r^4+9 (\ell^2+\ell-2)^3 (\ell (\ell+1) (\ell^2+\ell\nonumber\\
&+& 14)+144) r^5 \Big]
\eeq

Re-write Eq.~\eqref{eq:epsilon1mastereq} in the form
\beq
&& \frac{d^2\Psi_\pm^{\epsilon_1}}{dr_*^2}+\bigg(\omega^2 \left(1 + \epsilon_1 \sqrt{f_t^{\epsilon_1}f_r^{\epsilon_1}} \frac{16128 M^8}{r^8} \right) 
\nonumber \\ 
&&-\sqrt{f_t^{\epsilon_1}f_r^{\epsilon_1}}\Big(V_\pm^{\rm GR}
+\epsilon_1 \tilde{V}_\pm^{\epsilon_1}\Big)\bigg)\Psi_\pm^{\epsilon_1} =0\,.
\nonumber
\eeq
The potential $V_\pm^{\rm GR} +\epsilon_1 \tilde{V}_\pm^{\epsilon_1}$ is frequency-independent, and positive-definite outside the horizon for small positive $\epsilon_1$ (other arguments imply that $\epsilon_1,\epsilon_2$ are positive~\cite{Endlich:2017tqa}). It is possible to change $r_*$ coordinate and to transform the master equation into a manifestly positive-definite potential. Thus, mathematically, one can show mode stability of the above equation for {\it any} positive $\epsilon_1$~\cite{Kodama:2003kk}. Notice however that on physical grounds, large enough couplings (that might change the positive-definite character of the GR potential) are probably outside the regime of validity of the EFT.

\subsection{$\epsilon_2$ corrections}
Set now $\epsilon_1=\epsilon_3=0$. Then $f_t^{\epsilon_2} = f_r^{\epsilon_2} = f = 1- 2M/r$,
and the master equations take the form
\be
\frac{d^2\Psi_\pm^{\epsilon_2}}{dr_*^2}+\left(\omega^2-f \left(V_\pm^{\rm GR}+\epsilon_2 V_\pm^{\epsilon_2}\right)\right)\Psi_\pm^{\epsilon_2} =0\,,
\ee
where $dr/dr_*=f$. Defining functions $\Psi_\pm^{\epsilon_2}$
\beq
&& \Psi_-^{\epsilon_2} = \frac{i f h_1(r)}{r \omega }\,,\\
&& \Psi_+^{\epsilon_2} = \frac{1}{(\ell^2+\ell-2) r+6 M}\left[-r^2 K+\frac{i f r H_1}{\omega }\right]\,,
\eeq
we find 
\beq
V_-^{\epsilon_2} &=& \frac{4608M^8(\ell-1)\ell(\ell+1)(\ell+2)}{r^{10}}\,,\\
V_+^{\epsilon_2} &=& 0\,.
\eeq
The equation for even parity is the same as in GR. Since $V_-^{\epsilon_2}$ is clearly positive definite the system is linearly mode stable for positive $\epsilon_2$ as long as higher order corrections are negligible.

\subsection{$\epsilon_3$ corrections}
For $\epsilon_1=\epsilon_2=0$ we find that $\epsilon_3$ couples odd and even modes together.
Defining the master variables
\beq
\tilde{\Psi}_-^{\epsilon_3} &=& \Psi_-^{\epsilon_3} +\epsilon_3 \frac{64 M^8  \Psi_+^{\epsilon_3}}{r^9 ((\ell^2+\ell-2) r+6 M)}(-63 (\ell^2+\ell-2) r^2\nonumber\\ 
&+&2 (56 \ell (\ell+1)-355) M r+888 M^2)\nonumber\\ 
&+& \epsilon_3 \frac{1152 M^8}{r^7}\frac{d\Psi_-^{\epsilon_3}}{dr_*}\,,\\
\tilde{\Psi}_+^{\epsilon_3}&=& \Psi_+^{\epsilon_3} +\epsilon_3 \frac{64 M^8 \Psi_-^{\epsilon_3}}{r^9 ((\ell^2+\ell-2) r+6 M)}
(2 (70 \ell (\ell+1)\nonumber\\ 
&-& 383) M r+3 (\ell-1) (\ell+2) (2 \ell (\ell+1)-21) r^2
\nonumber\\ &+& 1056 M^2)-\epsilon_3 \frac{1152 M^8}{r^7} \frac{d\Psi_-^{\epsilon_3}}{dr_*}\,,
\eeq
with
\beq
&& \Psi_-^{\epsilon_3} = \frac{i f h_1(r)}{r \omega }\,,\\
&& \Psi_+^{\epsilon_3}  = \frac{1}{(\ell^2+\ell-2) r+6 M}\left[-r^2 K+\frac{i f r H_1}{\omega }\right]\,,
\eeq
the perturbed field equations can be written in the form
\beq
&& \frac{d^2\tilde{\Psi}_-^{\epsilon_3} }{dr_*^2}+(\omega^2- fV_-^{\rm GR})\tilde{\Psi}_-^{\epsilon_3}-\epsilon_3 f V^{\epsilon_3} \tilde{\Psi}_+^{\epsilon_3} =0\,,\\
&& \frac{d^2\tilde{\Psi}_+^{\epsilon_3} }{dr_*^2}+(\omega^2- fV_+^{\rm GR})\tilde{\Psi}_+^{\epsilon_3}-\epsilon_3 f V^{\epsilon_3} \tilde{\Psi}_-^{\epsilon_3} =0\,,
\eeq
where $dr/dr_*=f$, and $V^{\epsilon_3}$ is given by
\beq
&& V^{\epsilon_3} = -\frac{384 M^8}{r^{12} ((\ell^2+\ell-2) r+6 M)^2}\nonumber\\ 
&& \times\Big[ 203280 M^4+ 2400 (31 \ell (\ell+1)-132) M^3 r\nonumber\\ 
&& +4 (\ell (\ell+1) (1743 \ell (\ell+1)-22394)+46321) M^2 r^2\nonumber\\
&& + 36 (\ell-1) (\ell+2) (\ell (\ell+1) (\ell^2+\ell-162)+671) M r^3\nonumber\\ 
&& + 3 (\ell^2+\ell-2)^2 ((\ell-1) \ell (\ell+1) (\ell+2)+396) r^4\Big].
\eeq
It is easy to check that $V^{\epsilon_3}/(V_-^{\rm GR}-V_+^{\rm GR})$ is not a constant,
and hence decoupling of the above equations is not possible~\cite{Chuan:1992}.

We can see that $\epsilon_3$ appears only in terms coupling even and odd modes, i.e., non-diagonal parts of the equations.
For the potential matrix 
\be
\bm{V}_{(\epsilon_3)} = 
f 
\begin{pmatrix}
V_-^{\rm GR} & \epsilon_3 V^{\epsilon_3} \\
\epsilon_3 V^{\epsilon_3} & V_+^{\rm GR}\,.
\end{pmatrix}
\ee
the determinant, $\det(\bm{V}_{(\epsilon_3)}) = f^2 V_-^{\rm GR}V_+^{\rm GR} + {\cal O}(\epsilon_3^2)$,
is positive in $r>2M$. It is easy to check that both eigenvalues of $\bm{V}_{(\epsilon_3)}$ are positive in the far region. 
Thus, following the argument in Ref.~\cite{Kimura:2018nxk}, mathematically this system is mode-stable
for any coupling (within the small coupling regime of the effective field theory).

The $\epsilon_3$ parameter couples odd and even modes, and introduces non-trivial effects in the tidal deformability. Consider static, quadrupolar ($\ell = 2$) perturbations,
for which the general solution regular at the horizon is
\beq
H_0 &=&-r^2 (1-2 M/r) {\cal E}_2-\frac{384 M^5 {\cal B}_2 \epsilon_3}{35 r^8}\times \Big(420 M^5\nonumber \\ 
&-& 280 M^4 r-135 M^3 r^2+50 M^2 r^3+ 21 M r^4+7 r^5\Big)\,,\nonumber\\
h_0 &=&\frac{1}{3} r^3 {\cal B}_2 (1-2 M/r)+\frac{64 M^5 {\cal E}_2 \epsilon_3 (r- 2M)}{105 r^7}\nonumber \\ 
&\times & (980 M^4+630 M^3 r-540 M^2 r^2-210 M r^3-63 r^4)\,,
\nonumber
\eeq
where ${\cal E}_2,\,{\cal B}_2$ are the strength of the external gravitational quadrupolar moment, polar and axial, respectively.
Thus, an external {\it axial} quadrupolar field, induces a {\it polar} quadrupolar moment in the spacetime.
New TLNs seem to be needed to describe this scenario, but we will not dwell on this any further.

\section{Higher order correction to the spherically symmetric solution}
The spherically symmetric solution is described by Eqs.~\eqref{gtt1}-\eqref{grr2}, up to terms which are linear in
the coupling constant $\epsilon_1$. Some of the higher-order corrections can be computed directly from the action,
taking $f_{t,r}^{\epsilon_1} = 1 - 2 M/r + \epsilon_1 \delta f_{t,r}^{\epsilon_1} + \epsilon_1^2 \delta f_{t,r}^{\epsilon_1^2}$, 
where $\delta f_{t,r}^{\epsilon_1}$ are given by Eqs.~\eqref{gtt1}-\eqref{grr2}.
Expanding the action up to quadratic order in $f_{t,r}^{\epsilon_1^2}$, i.e., ${\cal O}(\epsilon_1^4)$, we can vary the action with respect to these two unknown functions. 
We obtain the asymptotically flat solution,
\beq
f_{t}^{\epsilon_1^2} &=&  
-\frac{16384 M^{17} }{17 r^{19}}(198305 M^2-231709 M r+64584 r^2),\nonumber\\
f_{r}^{\epsilon_1^2} &=&  -\frac{16384 M^{17} }{17 r^{19}}(2513399 M^2-2370222 M r+554472 r^2)\nonumber\,.
\eeq

Notice that nonlinear terms fall-off much faster than the linear corrections.
We can now also roughly estimate the range of validity of this solution, by requiring that $\delta f_{t,r}^{\epsilon_1^2}<\delta f_{t,r}^{\epsilon_1}$
everywhere in the BH exterior. We find $\epsilon_1\lesssim 340/9157$. However, one should realize that in the EFT of Ref.~\cite{Endlich:2017tqa}, the leading correction comes from operators schematically of the form $\left(R_{\mu\nu\rho\sigma}{}\right)^5$ or $\nabla^2 \left(R_{\mu\nu\rho\sigma}{}^4\right)$. These terms give rise to corrections of order $1/\Lambda^8\sim \epsilon^{4/3}$, which are much larger the one we consider. In other words, while the leading effect is of order $\epsilon$, subleading corrections are suppressed by additional powers of $\epsilon^{1/3}$.

\section{${\cal C}^n$ theory}
Consider a theory with the action
\beq
S_{{\cal{C}}^n} = \int d^4 x \sqrt{-g}\,2 M_{\rm pl}^2 \left(R - \frac{1}{\bar{\Lambda}^{2(2n-1)}}  {\cal C}^n\right)\,,
\label{eq:eftactioncn}
\eeq
and look for geometries which are small deviations from the Schwarzschild spacetime,
$f_t=1-2M/r+\alpha_n \delta f_t$ and $f_r=1-2 M/r+\alpha_n \delta f_r$,
where $\alpha_n = 1/(M\bar{\Lambda})^{2(2n-1)}$. We can use this ansatz in Eq.~\eqref{eq:eftactioncn}, expand it 
up to second order of $\delta f_t$ and $\delta f_r$ and take the variation with respect to $\delta f_t$ and $\delta f_r$. We find
the equation
\beq
&& (r \delta f_r)^\prime = 2^{4 n-1} 3^n (n-1) M^{6 n-3} r^{2-6 n} \nonumber \\ 
&& \times  \Big((2 (7-12 n) n+1)M +4 n (3 n-2) r\Big), \\
&& \bigg( \bigg(1 - \frac{2M}{r}\bigg)^{-1}\delta f_t\bigg)^\prime +\frac{r \delta f_r}{(r-2 M)^2} 
=\frac{M^{6 n-3} r^{2-6 n} }{r-2 M}\nonumber \\
&& \times 2^{4 n-1} 3^n (n-1)   (-6 M n+M+2 n r).
\eeq
We find the asymptotic flat solutions as
\beq
\delta f_t &=& 
-\frac{48^n (n-1) (M (1-6 n)+4 n r)}{6 M (2 n-1)}\left(\frac{M}{r}\right)^{6 n-2},
\\
\delta f_r &=&
-\frac{48^n (n-1) }{6 M (2 n-1)}\left(\frac{M}{r}\right)^{6 n-2}
\nonumber \\ 
& \times &
(M (2 (7-12 n) n+1)+6 n (2 n-1) r).
\eeq
The location of the horizon becomes
\be
r = r_H = 2 M \left(1 + 
\alpha_n 
\frac{3^{n-1} (2 n^2-n-1)}{2^{2n-1}(2 n-1)}
  \right).
\ee

\bibliographystyle{h-physrev4}
\bibliography{EFT_refs}

\begin{thebibliography}{10}
\bibitem{EFT}
All EFTs have some numerical ambiguity in defining the expansion parameter. The
  actual definition is inconsequential for the correctness of the calculation,
  but it indicates the range of validity of the framework and the actual size
  of the corrections. With our choice, the corrections to the metric are of
  order $\epsilon$ at $r\sim r_H$, a reasonable outcome. A more sensible
  definition would consider the magnitude of the next-order correction, but we
  leave this improved definition of $\epsilon$ to future work.
\bibitem{units}
The seemingly large coefficients in the metric corrections are only artifacts
  of our choice of parametrization: had we chosen $r_h$ instead of the mass
  $m$, all coefficients would be of order unity.
\end{thebibliography}

\end{document}